# A new multivariate and non-parametric association measure based on paired orthants

Eloi Martinez-Rabert[1,2]

[1]Department of Biochemistry, University of Wisconsin-Madison, Madison, WI, USA

[2]James Watt School of Engineering, Infrastructure and Environment Research Division, University of Glasgow, Advanced Research Centre, United Kingdom

## Abstract

Multivariate correlation analysis plays a key role in various fields such as statistics and big data analytics. In this paper, it is presented a new non-parametric association measure between more than two variables based on the concept of *paired orthants*. In order to evaluate the proposed methodology, different *N*-tuple sets (from two to six variables) have been evaluated. The presented rank correlation analysis not only evaluates the inter-relatedness of multiple variables, but also determine the specific tendency of these variables.

# 1. Introduction

Multivariate correlation analysis would play an important role in ecology for the evaluation and interpretation of interactions between more than two species (known as higher-order interactions) (Ludington, 2022). Among the existing correlations analyses for two variables, Kendall's $\tau$ correlation (Kendall, 1938) is of special interest because (*i*) is based on either being distribution-free or having a specified distribution but without specific parameters (i.e., a non-parametric measure), (*ii*) the simplicity of its methodology and (*iii*) the property of $\tau$ to be close to a normality distribution even for low number of observations (*n*). For two observed variables (*X* and *Y*), the Kendall's $\tau$ correlation is defined as

$$\tau = \frac{(number\ of\ concordant\ pairs) - (number\ of\ discordant\ pairs)}{(number\ of\ pairs)}.$$

For a multivariate case (i.e., for observations of more than two variables, $N \geq 2$), the average of the pairwise linear correlations is widely used as a measure of multivariate dependence (Capiello & Engle; Joe, 1989; Kendall & Smith, 1940; Moskowitz, 2015; Pollet & Wilson, 2010). Other approaches have been proposed for the measure of multivariate dependence based on Kendall's $\tau$, such as sign function-based multivariate extension (Simon, 1977), estimation based on multivariate cumulative distribution functions and multivariate survival functions (Joe, 1990), conditional copula-based extension (Ascorbebeitia et al., 2022) or absolute distance-based extension (Deng et al., 2022). All these are methods for assessing the degree of *concordance* between multiple variables.

In this paper, a new multivariate and non-parametric method that describes the association between more than two variables is proposed. This methodology can be employed to state which one is the significant tendency of multiple joint variables, beyond the concordance between them, allowing the description of complex data trends such as the aforementioned higher-order interactions. Due to the loss of information associated with the concept of *discordance* when more than two observed variables are considered (addressed in the following section), this methodology has been developed based on the concept of *paired orthants*.

# 2. Limitations of the *discordance* concept for multiple variables

The intuitive way to define the *concordance* and *discordance* concepts is this: two random variables (*X* and *Y*) are *concordant* when large (or low) values of *X* go with large (or low) values of *Y*, whereas these two are *discordant* when large values of *X* go with low values of *Y* or vice versa.

Let $(x_1, y_1), \ldots, (x_n, y_n)$ be a set of observations of the joint random variables $X$ and $Y$, such that all the values of $(x_i)$ and $(y_i)$ are unique. If a reference point is fixed $((x_i^*, y_i^*) \in \mathbb{R}^2)$, it is possible to determine four subsets of $\mathbb{R}^2$ (quadrants, $Q$):

$$Q_{\text{I}}(x_i^*, y_i^*) = \{(x_j, y_j) \in \mathbb{R}^2 : x_i^* < x_j, y_i^* < y_j\}$$

$$Q_{\text{II}}(x_i^*, y_i^*) = \{(x_j, y_j) \in \mathbb{R}^2 : x_i^* > x_j, y_i^* < y_j\}$$

$$Q_{\text{III}}(x_i^*, y_i^*) = \{(x_j, y_j) \in \mathbb{R}^2 : x_i^* > x_j, y_i^* > y_j\}$$

$$Q_{\text{IV}}(x_i^*, y_i^*) = \{(x_j, y_j) \in \mathbb{R}^2 : x_i^* < x_j, y_i^* > y_j\}$$

**Definition 1.** *Let $(x_i^*, y_i^*, \ldots, k_i^*)$ be the reference point of a set of observations of K joint random variables $\langle X, Y, \ldots, K \rangle$. An orthant in N-dimensions is considered the intersection of mutually orthogonal half-spaces through the reference point. An orthant in 2-dimensions is a quadrant (Q), in 3-dimensions an octant (O), and for more than three a hyperoctant (nH).*

Note that on the quadrant $Q_{\text{I}}$ ($Q_{\text{III}}$) large (small) values of $X$ go with large (small) values of $Y$ – definition of *concordance*; and on the quadrant $Q_{\text{II}}$ ($Q_{\text{IV}}$) small (large) values of $X$ go with large (small) values of $Y$ – definition of *discordance* (Figure 1).

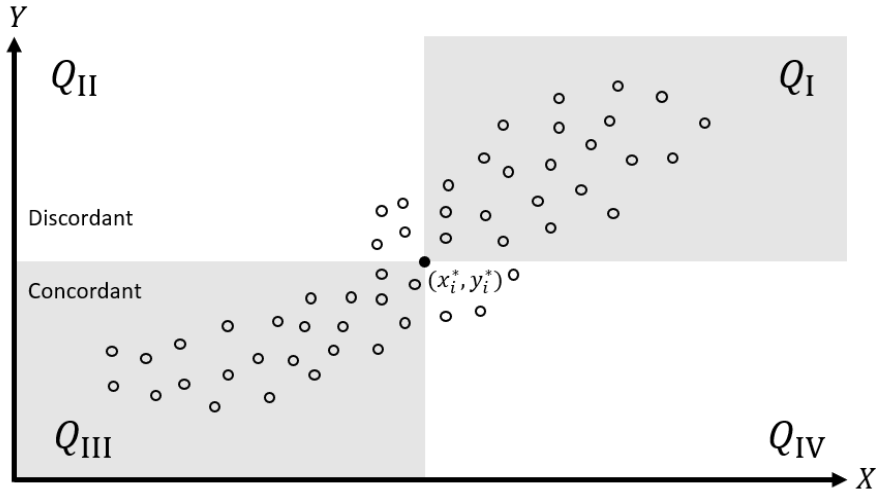

**Figure 1. Visual representation of *concordance* and *discordance* for two joint variables ($X$ and $Y$).** All points in the grey area are *concordant*, and all points in the white area are *discordant* with respect to $(x_i^*, y_i^*)$.

**Definition 2.** *Let H and H′ two independent set of observations. H is more concordant than H′ if, for all $(x, y) \in \mathbb{R}^2$,*

$$\Pr\{(x, y) \in [Q_{\text{I}}(x, y) \vee Q_{\text{III}}(x, y)] \mid H\} > \Pr\{(x, y) \in [Q_{\text{I}}(x, y) \vee Q_{\text{III}}(x, y)] \mid H'\}$$

*or H is less discordant than H′ if, for all $(x, y) \in \mathbb{R}^2$,*

$$\Pr\{(x, y) \in [Q_{\text{II}}(x, y) \vee Q_{\text{IV}}(x, y)] \mid H\} < \Pr\{(x, y) \in [Q_{\text{II}}(x, y) \vee Q_{\text{IV}}(x, y)] \mid H'\}.$$

Existing non-parametric rank correlation analysis between two random variables, such as Kendall's τ (Kendall, 1938) or Spearman's ρ (Spearman, 1904), are based on the probabilities of *concordance* and *discordance*. These are called *measures of concordance* because satisfy the set of axioms propose by Scarsini (Scarsini, 1984). Considering the Scarsini's axioms, the following result is immediate for a pair of random variables:

**Preposition 1.** *A pair of observations, $(x_i, y_i)$ and $(x_j, y_j)$, are concordant if have the property $sgn(x_j - x_i) = sgn(y_j - y_i)$. They are discordant if satisfy $sgn(x_j - x_i) = -sgn(y_j - y_i)$ (or $sgn(x_j - x_i) \neq sgn(y_j - y_i)$). $sgn(\chi)$ is the signum function of $\chi$:*

$$\mathrm{sgn}(\chi) := \begin{cases} -1 & if\ \chi < 0, \\ 0 & if\ \chi = 0, \\ 1 & if\ \chi > 0. \end{cases} \tag{1}$$

Now the ordinary statement of *concordance* and *discordance* is defined by three variables (*X*, *Y* and *Z*), the maximum set in which can be clearly represented with the Cartesian coordinate system. Let $(x_1, y_1, z_1), \ldots, (x_n, y_n, z_n)$ be a set of observations of the joint random variables *X*, *Y* and *Z*, such that all the values of $(x_i)$, $(y_i)$ and $(z_i)$ are unique. If a reference point $(x_i^*, y_i^*, z_i^*) \in \mathbb{R}^3$ is fixed, it is possible to determine eight subsets of $\mathbb{R}^3$ (octants, $O$):

$$O_\mathrm{I}(x_i^*, y_i^*, z_i^*) = \{(x_j, y_j, z_j) \in \mathbb{R}^3 : x_i^* < x_j, y_i^* < y_j, z_i^* < z_j\}$$

$$O_\mathrm{II}(x_i^*, y_i^*, z_i^*) = \{(x_j, y_j, z_j) \in \mathbb{R}^3 : x_i^* < x_j, y_i^* < y_j, z_i^* > z_j\}$$

$$O_\mathrm{III}(x_i^*, y_i^*, z_i^*) = \{(x_j, y_j, z_j) \in \mathbb{R}^3 : x_i^* > x_j, y_i^* < y_j, z_i^* > z_j\}$$

$$O_\mathrm{IV}(x_i^*, y_i^*, z_i^*) = \{(x_j, y_j, z_j) \in \mathbb{R}^3 : x_i^* > x_j, y_i^* < y_j, z_i^* < z_j\}$$

$$O_\mathrm{V}(x_i^*, y_i^*, z_i^*) = \{(x_j, y_j, z_j) \in \mathbb{R}^3 : x_i^* < x_j, y_i^* > y_j, z_i^* < z_j\}$$

$$O_\mathrm{VI}(x_i^*, y_i^*, z_i^*) = \{(x_j, y_j, z_j) \in \mathbb{R}^3 : x_i^* < x_j, y_i^* > y_j, z_i^* > z_j\}$$

$$O_\mathrm{VII}(x_i^*, y_i^*, z_i^*) = \{(x_j, y_j, z_j) \in \mathbb{R}^3 : x_i^* > x_j, y_i^* > y_j, z_i^* > z_j\}$$

$$O_\mathrm{VIII}(x_i^*, y_i^*, z_i^*) = \{(x_j, y_j, z_j) \in \mathbb{R}^3 : x_i^* > x_j, y_i^* > y_j, z_i^* < z_j\}$$

Note that on the octant $O_\mathrm{I}$ ($O_\mathrm{VII}$) large (small) values of *X* go with large (small) values of *Y* and *Z* – definition of *concordance*; whereas on the octant $O_\mathrm{II}$ ($O_\mathrm{VIII}$) large (small) values of *X* go with large (small) values of *Y* and small (large) values of *Z*, octant $O_\mathrm{III}$ ($O_\mathrm{V}$) small (large) values of *X* go with large (small) values of *Y* and small (large) values of *Z*, and octant $O_\mathrm{IV}$ ($O_\mathrm{VI}$) small (large) values of *X* go with large (small) values of *Y* and large (small) values of *Z* – all of them do not follow the *concordance* definition and, therefore, would be interpreted as definitions of *discordance* (Figure 2).

Then, for any pair of observations $(x_i, y_i, ..., k_i)$ and $(x_j, y_j, ..., k_j)$ from a set of $K$ variables $\langle X, Y, Z, ..., K \rangle$, the following result is direct:

**Preposition 2.** *A pair of observations, each with K variables, $(x_i, y_i, ..., k_i)$ and $(x_j, y_j, ..., k_j)$ are concordant if have the property $sgn(x_j - x_i) = sgn(y_j - y_i) = \cdots = sgn(k_j - k_i)$. Contrary, they are discordant if (at least) one non-equal is found (i.e., $... = sgn(z_j - z_i) \neq sgn(c_j - c_i) = \cdots$).*

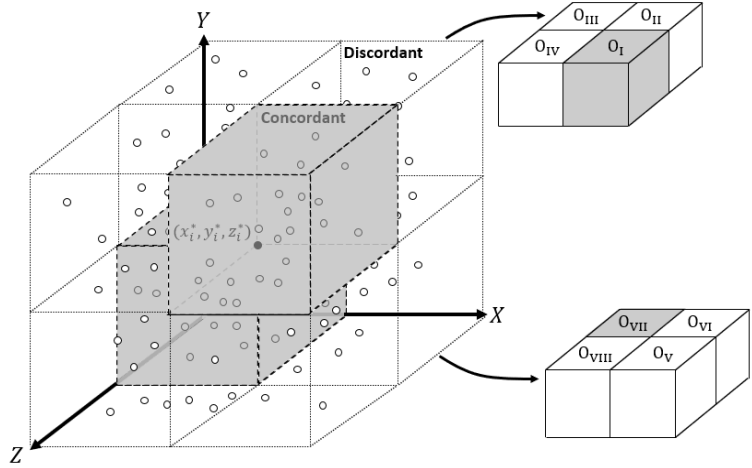

**Figure 2.** Representation of *concordance* and *discordance* for three joint variables $(X, Y, Z)$. All points in grey region are *concordant*, and all points in white region are *discordant* with respect to $(x_i^*, y_i^*, z_i^*)$.

The binary nature of *concordance*/*discordance* condition (i.e., if data set is not *concordant* then is *discordant* and vice versa) implies a loss of information in *discordance* definition when more than two observed variables are considered ($H(x, y, ..., k) \in \mathbb{R}^N$). For example, in the set of three observed variables ($H(x, y, z) \in \mathbb{R}^3$), these situations

*(i)* large (small) values of $X$ go with large (small) values of $Y$ and small (large) values of $Z$,
*(ii)* small (large) values of $X$ go with large (small) values of $Y$ and small (large) values of $Z$,
*(iii)* small (large) values of $X$ go with large (small) values of $Y$ and large (small) values of $Z$,

are not the same, although all of them would be considered *discordant* situations. Each of them is differentiable to the others, having its own *data trend*.

## 3. Definition of *paired orthants* and individual *data trends* coefficients

The orthants of a $N$-dimension ($\mathbb{R}^N$) can be associated to a set of signs ($[\pm]$). Let $(x_i^*, y_i^*, ..., k_i^*)$ the reference point of a set of observations of the joint random variables $\langle X, Y, ..., K \rangle$. The associated signs of each orthant is set based on the difference between the dataset $(x_j, y_j, ..., k_j)$ and the reference point (i.e., $\Delta k_{ji}^* = k_j - k_i^*$). For example, for a set $H(x, y) \in \mathbb{R}^2$, the associated signs for each quadrant ($Q$) are:

$$Q_I(x_i^*, y_i^*) = \{(x_j, y_j) \in \mathbb{R}^2 : \Delta x_{ji}^* > 0, \Delta y_{ji}^* > 0\} \equiv [+,+]$$

$$Q_{II}(x_i^*, y_i^*) = \{(x_j, y_j) \in \mathbb{R}^2 : \Delta x_{ji}^* < 0, \Delta y_{ji}^* > 0\} \equiv [-,+]$$

$$Q_{III}(x_i^*, y_i^*) = \{(x_j, y_j) \in \mathbb{R}^2 : \Delta x_{ji}^* < 0, \Delta y_{ji}^* < 0\} \equiv [-,-]$$

$$Q_{IV}(x_i^*, y_i^*) = \{(x_j, y_j) \in \mathbb{R}^2 : \Delta x_{ji}^* > 0, \Delta y_{ji}^* < 0\} \equiv [+,-]$$

and the associated signs of each octant ($O$) are:

$$O_I(x_i^*, y_i^*, z_i^*) = \{(x_j, y_j, z_j) \in \mathbb{R}^3 : \Delta x_{ji}^* > 0, \Delta y_{ji}^* > 0, \Delta z_{ji}^* > 0\} \equiv [+,+,+]$$

$$O_{II}(x_i^*, y_i^*, z_i^*) = \{(x_j, y_j, z_j) \in \mathbb{R}^3 : \Delta x_{ji}^* > 0, \Delta y_{ji}^* > 0, \Delta z_{ji}^* < 0\} \equiv [+,+,-]$$

$$O_{III}(x_i^*, y_i^*, z_i^*) = \{(x_j, y_j, z_j) \in \mathbb{R}^3 : \Delta x_{ji}^* < 0, \Delta y_{ji}^* > 0, \Delta z_{ji}^* < 0\} \equiv [-,+,-]$$

$$O_{IV}(x_i^*, y_i^*, z_i^*) = \{(x_j, y_j, z_j) \in \mathbb{R}^3 : \Delta x_{ji}^* < 0, \Delta y_{ji}^* > 0, \Delta z_{ji}^* > 0\} \equiv [-,+,+]$$

$$O_V(x_i^*, y_i^*, z_i^*) = \{(x_j, y_j, z_j) \in \mathbb{R}^3 : \Delta x_{ji}^* > 0, \Delta y_{ji}^* < 0, \Delta z_{ji}^* > 0\} \equiv [+,-,+]$$

$$O_{VI}(x_i^*, y_i^*, z_i^*) = \{(x_j, y_j, z_j) \in \mathbb{R}^3 : \Delta x_{ji}^* > 0, \Delta y_{ji}^* < 0, \Delta z_{ji}^* < 0\} \equiv [+,-,-]$$

$$O_{VII}(x_i^*, y_i^*, z_i^*) = \{(x_j, y_j, z_j) \in \mathbb{R}^3 : \Delta x_{ji}^* < 0, \Delta y_{ji}^* < 0, \Delta z_{ji}^* < 0\} \equiv [-,-,-]$$

$$O_{VIII}(x_i^*, y_i^*, z_i^*) = \{(x_j, y_j, z_j) \in \mathbb{R}^3 : \Delta x_{ji}^* < 0, \Delta y_{ji}^* < 0, \Delta z_{ji}^* > 0\} \equiv [-,-,+]$$

The orthants determined for $\mathbb{R}^2$ (quadrants, $Q$) and $\mathbb{R}^3$ (octants, $O$) can be *paired* based on their associated signs (pairing these through the equality: $Oh_\alpha = \neg Oh_\beta$. For an observation with two joint variables (H$(x,y) \in \mathbb{R}^2$), the *paired quadrants* ($Q_\alpha \oslash Q_\beta$) are:

$$Q_I(x_i^*, y_i^*) = \neg Q_{III}(x_i^*, y_i^*) \equiv [+,+] = \neg[-,-] \implies Q_I \oslash Q_{III}$$

$$Q_{II}(x_i^*, y_i^*) = \neg Q_{VI}(x_i^*, y_i^*) \equiv [-,+] = \neg[+,-] \implies Q_{II} \oslash Q_{VI},$$

where $Q_I \oslash Q_{III}$ and $Q_{II} \oslash Q_{VI}$ would correspond to the *concordance* and *discordance* definition for a set H$(x,y) \in \mathbb{R}^2$, respectively (Figure 1).

For an observation with three joint variables (H$(x,y,z) \in \mathbb{R}^3$), the *paired octants* ($O_\alpha \oslash O_\beta$) are:

$$O_I(x_i^*, y_i^*, z_i^*) = \neg Q_{VII}(x_i^*, y_i^*, z_i^*) \equiv [+,+,+] = \neg[-,-,-] \implies O_I \oslash O_{VII}$$

$$O_{II}(x_i^*, y_i^*, z_i^*) = \neg Q_{VIII}(x_i^*, y_i^*, z_i^*) \equiv [+,+,-] = \neg[-,-,+] \implies O_{II} \oslash O_{VIII}$$

$$O_{III}(x_i^*, y_i^*, z_i^*) = \neg Q_V(x_i^*, y_i^*, z_i^*) \equiv [-,+,-] = \neg[+,-,+] \implies O_{III} \oslash O_V$$

$$O_{IV}(x_i^*, y_i^*, z_i^*) = \neg Q_{VI}(x_i^*, y_i^*, z_i^*) \equiv [-,+,+] = \neg[+,-,-] \implies O_{IV} \oslash O_{VI},$$

where $O_I \oslash O_{VII}$ would correspond to the *concordance* definition for a set H$(x,y,z) \in \mathbb{R}^3$. Then, for any observation with $K$ joint variables (H$(x,y,...,k) \in \mathbb{R}^N$), the following result is immediate:

**Preposition 3.** *A couple of orthants ($Oh_\alpha$ and $Oh_\beta$) of a given reference point $(x_i^*, y_i^*, \ldots, k_i^*)$ are paired if and only if satisfies the property $Oh_\alpha(x_i^*, y_i^*, \ldots, k_i^*) = \neg Oh_\beta(x_i^*, y_i^*, \ldots, k_i^*)$. For any number of joint variables ($N$), the total of paired orthants ($Oh_\alpha \oslash Oh_\beta$) is equivalent to $2^N/2$.*

The set of symbols associated to each *paired orthants* ($[\pm, \pm, \ldots, \pm] = \neg[\mp, \mp, \ldots, \mp]$) is what defines the *data trend*. A straightforward way to evaluate the correlation between two variables is through probabilities (Kendall, 1938). Then, the *data trend* for a set of observations from $H(x, \ldots, k) \in \mathbb{R}^N$ can be stated as the probability of the data set to reside on the *paired orthants*. For example, the probabilities of a set $H(x, y) \in \mathbb{R}^2$ to reside in one of the *paired quadrants* ($Q_\alpha \oslash Q_\beta$), and following the respective *data trends* ($[\pm, \pm] \oslash [\mp, \mp]$) are:

$$\Pi_{Q_{III}}^{Q_I} \equiv \Pi_{[-,-]}^{[+,+]} = \Pr\{(x,y) \in [Q_I(x,y) \vee Q_{III}(x,y)]|H\} \equiv \Pr\{(x,y) \to [[+,+] \vee [-,-]] \mid H\}$$

$$\Pi_{Q_{IV}}^{Q_{II}} \equiv \Pi_{[+,-]}^{[-,+]} = \Pr\{(x,y) \in [Q_{II}(x,y) \vee Q_{IV}(x,y)]|H\} \equiv \Pr\{(x,y) \to [[-,+] \vee [+,-]] \mid H\}$$

For a set $H(x, y, z) \in \mathbb{R}^3$, the probabilities to reside on the *paired octants* ($O_\alpha \oslash O_\beta$) and follows the respective *data trends* ($[\pm, \pm, \pm] \oslash [\mp, \mp, \mp]$) are:

$$\Pi_{O_{VII}}^{O_I} \equiv \Pi_{[-,-,-]}^{[+,+,+]} = \Pr\{(x,y,z) \in [O_I(x,y,z) \vee O_{VII}(x,y,z)] \mid H\} \equiv \Pr\{(x,y,z) \to [[+,+,+] \vee [-,-,-]] \mid H\}$$

$$\Pi_{O_{VIII}}^{O_{II}} \equiv \Pi_{[-,-,+]}^{[+,+,-]} = \Pr\{(x,y,z) \in [O_{II}(x,y,z) \vee O_{VIII}(x,y,z)] \mid H\} \equiv \Pr\{(x,y,z) \to [[+,+,-] \vee [-,-,+]] \mid H\}$$

$$\Pi_{O_V}^{O_{III}} \equiv \Pi_{[+,-,+]}^{[-,+,-]} = \Pr\{(x,y,z) \in [O_{III}(x,y,z) \vee O_V(x,y,z)] \mid H\} \equiv \Pr\{(x,y,z) \to [[-,+,-] \vee [+,-,+]] \mid H\}$$

$$\Pi_{O_{VI}}^{O_{IV}} \equiv \Pi_{[+,-,-]}^{[-,+,+]} = \Pr\{(x,y,z) \in [O_{IV}(x,y,z) \vee O_{VI}(x,y,z)] \mid H\} \equiv \Pr\{(x,y,z) \to [[-,+,+] \vee [+,-,-]] \mid H\}$$

Then, any individual *data trend* coefficient ($\delta_{[\mp]}^{[\pm]}$) can be defined as

$$\delta_{[\mp]}^{[\pm]} = \Pi_{Oh_\beta}^{Oh_\alpha} \equiv \frac{(\# \text{ of pairs in } Oh_\alpha \oslash Oh_\beta)}{\binom{n}{2}} \equiv \Pi_{[\mp]}^{[\pm]} = \frac{(\# \text{ of pairs following } [\pm]) + (\# \text{ of pairs following } [\mp])}{\binom{n}{2}}$$

where $\binom{n}{2} = \frac{n \cdot (n-1)}{2}$ is the binomial coefficient for the number of ways to choose two items from $n$ observations. The $\delta_{[\mp]}^{[\pm]}$ value lies between 0 and 1 inclusive, taking 1 if and only if all the probability mass lies on the one of the *paired orthants* ($Oh_\alpha \oslash Oh_\beta$), that is, all data set follows certain *data trend* ($[\pm] \oslash [\mp]$). The value of 0 states the null probability of following the respective *data trend* ($[\pm] \oslash [\mp]$) of the coefficient $\delta_{[\mp]}^{[\pm]}$. If observed variables are independent, then all $\delta_{[\mp]}^{[\pm]}$ coefficients are equal to $2/2^N$ (where $N$ is the number of joint variables). The value of $2/2^N$ is referred here as the *reliable point* ($r_p$).

Recalling the definition of the associated symbols ($\Delta k_{ji} = k_j - k_i^*$) and *data trends* ($[\pm] \oslash [\mp]$), the explicit expression of the individual *data trend* coefficients for $H(x,y) \in \mathbb{R}^2$ ($\delta_{[\mp,\mp]}^{[\pm,\pm]}$) are:

$$\delta_{[-,-]}^{[+,+]} = \frac{2}{n \cdot (n-1)} \cdot \sum_{j>i} ([\Delta x_{ji} > 0 \wedge \Delta y_{ji} > 0] \vee [\Delta x_{ji} < 0 \wedge \Delta y_{ji} < 0])$$

$$\delta_{[+,-]}^{[-,+]} = \frac{2}{n \cdot (n-1)} \cdot \sum_{j>i} ([\Delta x_{ji} < 0 \wedge \Delta y_{ji} > 0] \vee [\Delta x_{ji} > 0 \wedge \Delta y_{ji} < 0]),$$

and for $H(x, y, z) \in \mathbb{R}^3$, the explicit expression of $\delta_{[\mp,\mp,\mp]}^{[\pm,\pm,\pm]}$ coefficients are:

$$\delta_{[-,-,-]}^{[+,+,+]} = \frac{2}{n \cdot (n-1)} \cdot \sum_{j>i} ([\Delta x_{ji} > 0 \wedge \Delta y_{ji} > 0 \wedge \Delta z_{ji} > 0] \vee [\Delta x_{ji} < 0 \wedge \Delta y_{ji} < 0 \wedge \Delta z_{ji} < 0])$$

$$\delta_{[-,-,+]}^{[+,+,-]} = \frac{2}{n \cdot (n-1)} \cdot \sum_{j>i} ([\Delta x_{ji} > 0 \wedge \Delta y_{ji} > 0 \wedge \Delta z_{ji} < 0] \vee [\Delta x_{ji} < 0 \wedge \Delta y_{ji} < 0 \wedge \Delta z_{ji} > 0])$$

$$\delta_{[+,-,+]}^{[-,+,-]} = \frac{2}{n \cdot (n-1)} \cdot \sum_{j>i} ([\Delta x_{ji} < 0 \wedge \Delta y_{ji} > 0 \wedge \Delta z_{ji} < 0] \vee [\Delta x_{ji} > 0 \wedge \Delta y_{ji} < 0 \wedge \Delta z_{ji} > 0])$$

$$\delta_{[+,-,-]}^{[-,+,+]} = \frac{2}{n \cdot (n-1)} \cdot \sum_{j>i} ([\Delta x_{ji} < 0 \wedge \Delta y_{ji} > 0 \wedge \Delta z_{ji} > 0] \vee [\Delta x_{ji} > 0 \wedge \Delta y_{ji} < 0 \wedge \Delta z_{ji} < 0])$$

Then, for any observation with $K$ joint variables ($H(x, y, \ldots, k) \in \mathbb{R}^N$), the explicit expressions for any $\delta_{[\mp,\mp,\ldots,\mp]}^{[\pm,\pm,\ldots,\pm]}$ coefficient is:

$$\delta_{[\mp,\mp,\ldots,\mp]}^{[\pm,\pm,\ldots,\pm]} = \frac{2}{n \cdot (n-1)} \cdot \sum_{j>i} ([\Delta x_{ji} \gtrless 0 \wedge \Delta y_{ji} \gtrless 0 \wedge \ldots \wedge \Delta k_{ji} \gtrless 0] \vee [\Delta x_{ji} \lessgtr 0 \wedge \Delta y_{ji} \lessgtr 0 \wedge \ldots \wedge \Delta k_{ji} \lessgtr 0]),$$

where a reduced form of this expression would be:

$$\delta_N = \frac{2}{n \cdot (n-1)} \cdot \sum_{j>i} (\wedge [\Delta c_{ji} \gtrless 0] \vee \wedge [\Delta c_{ji} \lessgtr 0]), \quad \Delta c_{ji} = \langle \Delta x_{ji}, \Delta y_{ji}, \ldots, \Delta k_{ji} \rangle \qquad (2)$$

## 4. Definition of global *data trend* coefficients

The $\delta_N$ coefficients presented above correspond to the individual probability of each possible *data trend*. However, it would be more convenient to express these in terms of global probability (global *data trend* coefficients, $\iota_N$). For any number of joint variables ($X, Y, \ldots, K$), a total of $2^N/2$ (corresponding to the number of *paired orthants*) $\iota_N$ coefficients are needed to describe all possible *data trends* $\langle \iota_i, \iota_j, \ldots, \iota_d \rangle$. In order to estimate any $\iota_N$ coefficient (e.g., $(\iota_N)_i$), first it is computed the differences between the individual *data trend* coefficient $\delta_i$ and the others individual coefficients $\langle \delta_j, \ldots, \delta_d \rangle$. Then, a modified geometric mean (able to consider both positive and negative values) is applied to the calculated differences (de la Cruz & Kreft, 2018):

$$\left\{\prod_{i=1}^{nc}(1+a_i)\right\}^{(1/nc)} - 1$$

Where $a_i$ is the difference between the individual *data trend* coefficients, and $nc$ the number of comparisons between the individual *data trends* coefficients ($2^N/2 - 1$).

For two observed variables ($H(x,y) \in \mathbb{R}^2$), only one $\iota$ coefficient is sufficient to describe all possible *data trends*, corresponding to the Kendall's $\tau$ coefficient (Kendall, 1938):

$$\iota_{[-,-]}^{[+,+]} = \delta_{[-,-]}^{[+,+]} - \delta_{[-,+]}^{[+,-]}$$

$$\iota_{[-,+]}^{[+,-]} = \delta_{[-,+]}^{[+,-]} - \delta_{[-,-]}^{[+,+]}$$

$$\iota_{[-,-]}^{[+,+]} = -\iota_{[-,+]}^{[+,-]} \equiv \tau \qquad (3)$$

Considering a set of observation with three joint variables ($H(x,y,z) \in \mathbb{R}^3$), the global probabilities of their *data trends* ($\iota_{[\mp,\mp,\mp]}^{[\pm,\pm,\pm]}$) are evaluated by:

$$\iota_{[-,-,-]}^{[+,+,+]} = \sqrt[3]{\left\{1 + \left(\delta_{[-,-,-]}^{[+,+,+]} - \delta_{[-,-,+]}^{[+,+,-]}\right)\right\} \cdot \left\{1 + \left(\delta_{[-,-,-]}^{[+,+,+]} - \delta_{[+,-,+]}^{[-,+,-]}\right)\right\} \cdot \left\{1 + \left(\delta_{[-,-,-]}^{[+,+,+]} - \delta_{[+,-,-]}^{[-,+,+]}\right)\right\}} - 1$$

$$\iota_{[-,-,+]}^{[+,+,-]} = \sqrt[3]{\left\{1 + \left(\delta_{[-,-,+]}^{[+,+,-]} - \delta_{[-,-,-]}^{[+,+,+]}\right)\right\} \cdot \left\{1 + \left(\delta_{[-,-,+]}^{[+,+,-]} - \delta_{[+,-,+]}^{[-,+,-]}\right)\right\} \cdot \left\{1 + \left(\delta_{[-,-,+]}^{[+,+,-]} - \delta_{[+,-,-]}^{[-,+,+]}\right)\right\}} - 1$$

$$\iota_{[+,-,+]}^{[-,+,-]} = \sqrt[3]{\left\{1 + \left(\delta_{[+,-,+]}^{[-,+,-]} - \delta_{[-,-,-]}^{[+,+,+]}\right)\right\} \cdot \left\{1 + \left(\delta_{[+,-,+]}^{[-,+,-]} - \delta_{[-,-,+]}^{[+,+,-]}\right)\right\} \cdot \left\{1 + \left(\delta_{[+,-,+]}^{[-,+,-]} - \delta_{[+,-,-]}^{[-,+,+]}\right)\right\}} - 1$$

$$\iota_{[+,-,-]}^{[-,+,+]} = \sqrt[3]{\left\{1 + \left(\delta_{[+,-,-]}^{[-,+,+]} - \delta_{[-,-,-]}^{[+,+,+]}\right)\right\} \cdot \left\{1 + \left(\delta_{[+,-,-]}^{[-,+,+]} - \delta_{[-,-,+]}^{[+,+,-]}\right)\right\} \cdot \left\{1 + \left(\delta_{[+,-,-]}^{[-,+,+]} - \delta_{[+,-,+]}^{[-,+,-]}\right)\right\}} - 1$$

For any number of joint variables ($H(x,y,\dots,k) \in \mathbb{R}^N$), the general expression for $(\iota_N)_i$ coefficients correspond to:

$$(\iota_N)_i = \left[\prod_{j \neq i}\{1 + (\delta_i - \delta_j)\}\right]^{\left(\frac{2^N}{2}-1\right)^{-1}} - 1 \qquad (4)$$

The $\iota_N$ coefficient lies between –1 and 1 inclusive, taking 1 if and only if all the probability mass lies on the specific *data trend* ([±] ⊘ [∓]). The value of –1 states the null probability of the data set to follow this specific *trend*, following another *data trend* entirely. If observed variables are independent, then all $\iota_{[\mp]}^{[\pm]}$ coefficients are equal to 0. All $\delta_i$ coefficients above the $r_p$ will have a positive $(\iota_N)_i$ coefficient, whereas those $\delta_i$ coefficients below the $r_p$ will have a negative $(\iota_N)_i$ coefficient. The $\delta_i$ coefficients equal to $r_p$ will have a $(\iota_N)_i$ equal 0.

## 5. Sampling distribution and statistical hypothesis for $\iota_N$ coefficients

To measure the significance of an observed *data trend*, it is necessary to know whether the value could have arisen by chance from a universe in which all the possible rankings of *n* objects occur an equal number of times. It is, therefore, necessary to consider the distribution of *trend*s in such universe. The distribution of $\iota$ tends to normality for large number of observations and surprisingly close to normality even for low number of observations. The fact that $\tau$ tends to normality even for low number of observations was proved by Kendall (1938).

Then, the distribution of $\iota_N$ (in which all existing *data trends* also would occur with the same frequency) also converges towards a normal distribution with a mean of 0 and a variance equal to:

$$\sigma_\iota^2 = \frac{2(2n+5)}{9n(n-1)}.$$

Where *n* is the number of observations. Therefore, the null hypothesis test can be performed by transforming $\iota_N$ into statistic $Z_\iota$ as:

$$Z_\iota = \frac{\iota_N}{\sigma_\iota} = \frac{\iota_N}{\sqrt{\frac{2(2n+5)}{9n(n-1)}}} \tag{5}$$

Thus, to evaluate whether a set of variables significantly follows a certain trend, one computes $Z_\iota$ and finds the cumulative probability for a standard normal distribution at $-|Z_\iota|$.

## 6. Algorithm

Now, the algorithm for the direct estimation of all $(\iota_N)_i$ coefficients ($i = 1, \ldots, 2^N/2$, being *N* the number of joint variables) is described in the following steps.

Step 1. Obtention of the *data trends* ([±] ⊘ [∓]) associated to each *paired orthant* ($Oh_\alpha$ ⊘ $Oh_\beta$). First performing the permutation with repetition of the positive (+) and negative (−) signs, and then *pairing* the permutation results based on the equality [±] = ¬[∓].

Step 2. Calculation of all individual *data trend* coefficient ($\delta_i$) with the Equation 2. The direct computation of the summation $\sum_{j>i}(\ldots)$ involves two nested iterations:

```
f = 0;
for j = 2:n
        for i = 1:j-1
                diff = dataset(j,:) – dataset(i,:);
                f = f + or(eq(sym_up, gt(diff, 0)), eq(sym_down, gt(diff, 0)));
        end
end
```

Where n is the number of observations, sym_up is the set of $[\pm]$, sym_up is the set of $[\mp]$, gt(A, B) determines if 'A is greater than B', and eq(A, B) determines if 'A and B are equal'.

Step 3. Calculation of global *data trend* coefficients $(\iota_N)_i$ with Equation 3 for $N = 2$, and Equation 4 for $N > 2$.

Step 4. Statistical analysis with the Equation 5 and the normal cumulative distribution function. To obtain the *p*-value for a two-sided test, the number from the normal cumulative distribution function is multiplied by two.

$$p = 2 * normcdf(-|Z_\iota|).$$

## 7. Examples

The application of the proposed multivariate association measure ($\iota_N$ coefficients) to different sets of joint variables have been included: two and three variables on Table 1 ($H(S1, S2) \in \mathbb{R}^2$ and $H(S1, S2, S3) \in \mathbb{R}^3$, respectively), four variables on Table 2 ($H(S1, S2, S3, S4) \in \mathbb{R}^4$), five joint variables on Table 3 ($H(S1, S2, S3, S4, S5) \in \mathbb{R}^5$), and six joint variables on Table 4 ($H(S1, S2, S3, S4, S5, S6) \in \mathbb{R}^6$).

## 8. Conclusions

In this paper, a new non-parametric association measure for multiple variables ($N \geq 2$) is presented – the multivariate $\iota$ coefficient ($\iota_N$). This analysis able to evaluate with a unique formula the probability of all associations between multiple set of variables, that is, their specific trends. In order to avoid the loss of information about the dependences between *N* random variables, *concordant* and *discordant* concepts was replaced by *paired orthants* and *data trend* concepts to describe these dependencies.

## Data availability

The source code to produce the results presented in this manuscript is available on a public GitHub repository at https://github.com/soundslikealloy/multivarcorr.

## Disclosure statement

The author report there are no competing interests to declare.

**Table 1.** Examples of *data trend* measurement ($\iota^{[\pm]}_{[\mp]}$) for $H(S1, S2) \in \mathbb{R}^2$ and $H(S1, S2, S3) \in \mathbb{R}^3$ ($r_p = 2/2^N$).

| $N$ ($r_p$) | Series ($n = 12$) | $\delta\ [0, 1]$ | $\iota\ [-1, 1]$ | $p$ |
|---|---|---|---|---|
| **2** ($r_p = $ **0.5000**) | **S1**: 1, 3, 5, 7, 9, 11, 13, 15, 17, 19, 21, 23<br>**S2**: 1, 3, 5, 7, 9, 11, 13, 15, 17, 19, 21, 23 | $\delta^{[+,+]}_{[-,-]} = 1.0000$<br>$\delta^{[-,+]}_{[+,-]} = 0.0000$ | $\iota \equiv \tau = 1.0000$ | <0.0001 |
| | **S1**: 1, 3, 9, 7, 15, 13, 21, 23, 5, 11, 17, 19<br>**S2**: 23, 21, 15, 17, 9, 11, 3, 1, 19, 13, 7, 5 | $\delta^{[+,+]}_{[-,-]} = 0.0000$<br>$\delta^{[-,+]}_{[+,-]} = 1.0000$ | $\iota \equiv \tau = -1.0000$ | <0.0001 |
| | **S1**: 1, 3, 5, 7, 9, 11, 13, 15, 17, 19, 21, 23<br>**S2**: 1, 5, 9, 11, 6, 13, 17, 12, 22, 20, 25, 23 | $\delta^{[+,+]}_{[-,-]} = 0.9091$<br>$\delta^{[-,+]}_{[+,-]} = 0.0909$ | $\iota \equiv \tau = 0.8182$ | 0.0002 |
| | **S1**: 5, 8, 9, 16, 12, 1, 14, 3, 15, 21, 13, 1<br>**S2**: 7, 15, 3, 20, 24, 23, 24, 20, 21, 5, 23, 15<br>*100% random distribution | $\delta^{[+,+]}_{[-,-]} = 0.4918$<br>$\delta^{[-,+]}_{[+,-]} = 0.5082$ | $\iota \equiv \tau = -0.0164$ | 0.9409 |
| **3** ($r_p = $ **0.2500**) | **S1**: 1, 3, 5, 7, 9, 11, 13, 15, 17, 19, 21, 23<br>**S2**: 1, 3, 5, 7, 9, 11, 13, 15, 17, 19, 21, 23<br>**S3**: 1, 3, 5, 7, 9, 11, 13, 15, 17, 19, 21, 23 | $\delta^{[+,+,+]}_{[-,-,-]} = 1.0000$<br>$\delta^{[-,+,+]}_{[+,-,-]} = 0.0000$<br>$\delta^{[+,-,+]}_{[-,+,-]} = 0.0000$<br>$\delta^{[-,-,+]}_{[+,+,-]} = 0.0000$ | $\iota^{[+,+,+]}_{[-,-,-]} = 1.0000$<br>$\iota^{[-,+,+]}_{[+,-,-]} = -1.0000$<br>$\iota^{[+,-,+]}_{[-,+,-]} = -1.0000$<br>$\iota^{[-,-,+]}_{[+,+,-]} = -1.0000$ | <0.0001<br><0.0001<br><0.0001<br><0.0001 |
| | **S1**: 1, 3, 5, 7, 9, 11, 13, 15, 17, 19, 21, 23<br>**S2**: 23, 21, 19, 17, 15, 13, 11, 9, 7, 5, 3, 1<br>**S3**: 1, 3, 5, 7, 9, 11, 13, 15, 17, 19, 21, 23 | $\delta^{[+,+,+]}_{[-,-,-]} = 0.000$<br>$\delta^{[-,+,+]}_{[+,-,-]} = 0.000$<br>$\delta^{[+,-,+]}_{[-,+,-]} = 1.000$<br>$\delta^{[-,-,+]}_{[+,+,-]} = 0.0000$ | $\iota^{[+,+,+]}_{[-,-,-]} = -1.0000$<br>$\iota^{[-,+,+]}_{[+,-,-]} = -1.0000$<br>$\iota^{[+,-,+]}_{[-,+,-]} = 1.0000$<br>$\iota^{[-,-,+]}_{[+,+,-]} = -1.0000$ | <0.0001<br><0.0001<br><0.0001<br><0.0001 |
| | **S1**: 1, 3, 5, 7, 9, 11, 13, 15, 17, 19, 21, 23<br>**S2**: 23, 21, 19, 17, 15, 13, 11, 9, 7, 5, 3, 1<br>**S3**: 23, 21, 19, 17, 15, 13, 11, 9, 7, 5, 3, 1 | $\delta^{[+,+,+]}_{[-,-,-]} = 0.0000$<br>$\delta^{[-,+,+]}_{[+,-,-]} = 1.0000$<br>$\delta^{[+,-,+]}_{[-,+,-]} = 0.0000$<br>$\delta^{[-,-,+]}_{[+,+,-]} = 0.0000$ | $\iota^{[+,+,+]}_{[-,-,-]} = -1.0000$<br>$\iota^{[-,+,+]}_{[+,-,-]} = 1.0000$<br>$\iota^{[+,-,+]}_{[-,+,-]} = -1.0000$<br>$\iota^{[-,-,+]}_{[+,+,-]} = -1.0000$ | <0.0001<br><0.0001<br><0.0001<br><0.0001 |
| | **S1**: 23, 12, 19, 12, 9, 5, 3, 1, 20, 7, 12, 17<br>**S2**: 2, 13, 5, 10, 15, 4, 20, 23, 3, 17, 11, 8<br>**S3**: 22, 11, 19, 15, 4, 8, 3, 1, 21, 6, 13, 17 | $\delta^{[+,+,+]}_{[-,-,-]} = 0.1111$<br>$\delta^{[-,+,+]}_{[+,-,-]} = 0.0000$<br>$\delta^{[+,-,+]}_{[-,+,-]} = 0.8889$<br>$\delta^{[-,-,+]}_{[+,+,-]} = 0.0000$ | $\iota^{[+,+,+]}_{[-,-,-]} = -0.3502$<br>$\iota^{[-,+,+]}_{[+,-,-]} = -0.5378$<br>$\iota^{[+,-,+]}_{[-,+,-]} = 0.8511$<br>$\iota^{[-,-,+]}_{[+,+,-]} = -0.5378$ | 0.1610<br>0.0192<br>0.0002<br>0.0192 |
| | **S1**: 19, 8, 6, 16, 7, 9, 13, 4, 14, 10, 11, 22<br>**S2**: 1, 4, 4, 7, 13, 5, 20, 12, 17, 23, 23, 19<br>**S3**: 16, 22, 1, 2, 22, 23, 11, 9, 2, 20, 10, 5<br>*100% random distribution | $\delta^{[+,+,+]}_{[-,-,-]} = 0.2419$<br>$\delta^{[-,+,+]}_{[+,-,-]} = 0.2742$<br>$\delta^{[+,-,+]}_{[-,+,-]} = 0.1613$<br>$\delta^{[-,-,+]}_{[+,+,-]} = 0.3226$ | $\iota^{[+,+,+]}_{[-,-,-]} = -0.0130$<br>$\iota^{[-,+,+]}_{[+,-,-]} = 0.0302$<br>$\iota^{[+,-,+]}_{[-,+,-]} = -0.1189$<br>$\iota^{[-,-,+]}_{[+,+,-]} = 0.0958$ | 0.9530<br>0.8915<br>0.5905<br>0.6647 |

**Table 2. Examples of *data trend* measurement ($\iota^{[\pm]}_{[\mp]}$) for $H(S1, S2, S3, S4) \in \mathbb{R}^4$ ($r_p = 2/2^N$).**

| $N$ ($r_p$) | Series ($n = 12$) | $\delta$ [0, 1] | $\iota$ [−1, 1] | $p$ |
|---|---|---|---|---|
| **4** ($r_p$ = 0.1250) | **S1**: 23, 21, 19, 17, 15, 13, 11, 9, 7, 5, 3, 1 <br> **S2**: 23, 21, 19, 17, 15, 13, 11, 9, 7, 5, 3, 1 <br> **S3**: 1, 3, 5, 7, 9, 11, 13, 15, 17, 19, 21, 23 <br> **S4**: 23, 21, 19, 17, 15, 13, 11, 9, 7, 5, 3, 1 | $\delta^{[+,+,+,+]}_{[-,-,-,-]} = 0.0000$ | $\iota^{[+,+,+,+]}_{[-,-,-,-]} = -1.0000$ | <0.0001 |
| | | $\delta^{[-,+,+,+]}_{[+,-,-,-]} = 0.0000$ | $\iota^{[-,+,+,+]}_{[+,-,-,-]} = -1.0000$ | <0.0001 |
| | | $\delta^{[+,-,+,+]}_{[-,+,-,-]} = 0.0000$ | $\iota^{[+,-,+,+]}_{[-,+,-,-]} = -1.0000$ | <0.0001 |
| | | $\delta^{[-,-,+,+]}_{[+,+,-,-]} = 0.0000$ | $\iota^{[-,-,+,+]}_{[+,+,-,-]} = -1.0000$ | <0.0001 |
| | | $\delta^{[+,+,-,+]}_{[-,-,+,-]} = 1.0000$ | $\iota^{[+,+,-,+]}_{[-,-,+,-]} = 1.0000$ | <0.0001 |
| | | $\delta^{[-,+,-,+]}_{[+,-,+,-]} = 0.0000$ | $\iota^{[-,+,-,+]}_{[+,-,+,-]} = -1.0000$ | <0.0001 |
| | | $\delta^{[+,-,-,+]}_{[-,+,+,-]} = 0.0000$ | $\iota^{[+,-,-,+]}_{[-,+,+,-]} = -1.0000$ | <0.0001 |
| | | $\delta^{[-,-,-,+]}_{[+,+,+,-]} = 0.0000$ | $\iota^{[-,-,-,+]}_{[+,+,+,-]} = -1.0000$ | <0.0001 |
| | **S1**: 12, 3, 6, 11, 10, 9, 21, 22, 4, 7, 23, 4 <br> **S2**: 15, 13, 1, 9, 19, 5, 9, 11, 8, 23, 18, 9 <br> **S3**: 10, 1, 2, 14, 18, 11, 12, 16, 7, 5, 13, 6 <br> **S4**: 6, 10, 17, 4, 18, 5, 22, 13, 15, 14, 2, 11 <br> *100% random distribution | $\delta^{[+,+,+,+]}_{[-,-,-,-]} = 0.1905$ | $\iota^{[+,+,+,+]}_{[-,-,-,-]} = 0.0712$ | 0.7473 |
| | | $\delta^{[-,+,+,+]}_{[+,-,-,-]} = 0.0635$ | $\iota^{[-,+,+,+]}_{[+,-,-,-]} = -0.0745$ | 0.7361 |
| | | $\delta^{[+,-,+,+]}_{[-,+,-,-]} = 0.1429$ | $\iota^{[+,-,+,+]}_{[-,+,-,-]} = 0.0163$ | 0.9413 |
| | | $\delta^{[-,-,+,+]}_{[+,+,-,-]} = 0.0635$ | $\iota^{[-,-,+,+]}_{[+,+,-,-]} = -0.0745$ | 0.7361 |
| | | $\delta^{[+,+,-,+]}_{[-,-,+,-]} = 0.0476$ | $\iota^{[+,+,-,+]}_{[-,-,+,-]} = -0.0925$ | 0.6755 |
| | | $\delta^{[-,+,-,+]}_{[+,-,+,-]} = 0.1429$ | $\iota^{[-,+,-,+]}_{[+,-,+,-]} = 0.0163$ | 0.9413 |
| | | $\delta^{[+,-,-,+]}_{[-,+,+,-]} = 0.0476$ | $\iota^{[+,-,-,+]}_{[-,+,+,-]} = -0.0925$ | 0.6755 |
| | | $\delta^{[-,-,-,+]}_{[+,+,+,-]} = 0.3016$ | $\iota^{[-,-,-,+]}_{[+,+,+,-]} = 0.2006$ | 0.3639 |
| | **S1**: 1, 3, 5, 7, 9, 11, 13, 15, 17, 19, 21, 23 <br> **S2**: 23, 21, 22, 17, 16, 13, 15, 9, 4, 5, 3, 2 <br> **S3**: 1, 6, 5, 7, 9, 25, 13, 2, 17, 22, 11, 23 <br> **S4**: 20, 23, 19, 17, 15, 9, 11, 10, 7, 6, 3, 5 | $\delta^{[+,+,+,+]}_{[-,-,-,-]} = 0.0000$ | $\iota^{[+,+,+,+]}_{[-,-,-,-]} = -0.2034$ | 0.3573 |
| | | $\delta^{[-,+,+,+]}_{[+,-,-,-]} = 0.1818$ | $\iota^{[-,+,+,+]}_{[+,-,-,-]} = 0.0197$ | 0.9291 |
| | | $\delta^{[+,-,+,+]}_{[-,+,-,-]} = 0.0303$ | $\iota^{[+,-,+,+]}_{[-,+,-,-]} = -0.1657$ | 0.4533 |
| | | $\delta^{[-,-,+,+]}_{[+,+,-,-]} = 0.0152$ | $\iota^{[-,-,+,+]}_{[+,+,-,-]} = -0.1845$ | 0.4037 |
| | | $\delta^{[+,+,-,+]}_{[-,-,+,-]} = 0.0152$ | $\iota^{[+,+,-,+]}_{[-,-,+,-]} = -0.1845$ | 0.4037 |
| | | $\delta^{[-,+,-,+]}_{[+,-,+,-]} = 0.7273$ | $\iota^{[-,+,-,+]}_{[+,-,+,-]} = 0.6872$ | 0.0019 |
| | | $\delta^{[+,-,-,+]}_{[-,+,+,-]} = 0.0152$ | $\iota^{[+,-,-,+]}_{[-,+,+,-]} = -0.1845$ | 0.4037 |
| | | $\delta^{[-,-,-,+]}_{[+,+,+,-]} = 0.0152$ | $\iota^{[-,-,-,+]}_{[+,+,+,-]} = -0.1845$ | 0.4037 |

**Table 3.** Examples of *data trend* measurement ($\iota_{[\mp]}^{[\pm]}$) for $H(S1, S2, S3, S4, S5) \in \mathbb{R}^5$ ($r_p = 2/2^N$).

| $N (r_p)$ | Series ($n = 12$) | $\delta [0, 1]$ | $\iota [-1, 1]$ | $p$ |
|---|---|---|---|---|
| **5** ($r_p = 0.0625$) | **S1**: 22, 14, 12, 7, 21, 11, 9, 23, 17, 6, 18, 20<br>**S2**: 9, 2, 18, 10, 16, 21, 7, 4, 17, 22, 19, 14<br>**S3**: 16, 21, 15, 5, 18, 13, 20, 7, 17, 2, 19, 9<br>**S4**: 6, 16, 15, 8, 5, 17, 10, 12, 3, 11, 4, 18<br>**S5**: 8, 2, 22, 19, 15, 11, 3, 23, 14, 7, 16, 21<br>*100% random distribution* | $\delta_{[-,-,-,-,-]}^{[+,+,+,+,+]} = 0.0455$ | $\iota_{[-,-,-,-,-]}^{[+,+,+,+,+]} = -0.0194$ | 0.9299 |
| | | $\delta_{[+,-,-,-,-]}^{[-,+,+,+,+]} = 0.0000$ | $\iota_{[+,-,-,-,-]}^{[-,+,+,+,+]} = -0.0679$ | 0.7587 |
| | | $\delta_{[-,+,-,-,-]}^{[+,-,+,+,+]} = 0.0909$ | $\iota_{[-,+,-,-,-]}^{[+,-,+,+,+]} = 0.0291$ | 0.8952 |
| | | $\delta_{[+,+,-,-,-]}^{[-,-,+,+,+]} = 0.0000$ | $\iota_{[+,+,-,-,-]}^{[-,-,+,+,+]} = -0.0679$ | 0.7587 |
| | | $\delta_{[-,-,+,-,-]}^{[+,+,-,+,+]} = 0.0606$ | $\iota_{[-,-,+,-,-]}^{[+,+,-,+,+]} = -0.0033$ | 0.9882 |
| | | $\delta_{[+,-,+,-,-]}^{[-,+,-,+,+]} = 0.1212$ | $\iota_{[+,-,+,-,-]}^{[-,+,-,+,+]} = 0.0616$ | 0.7806 |
| | | $\delta_{[-,+,+,-,-]}^{[+,-,-,+,+]} = 0.1212$ | $\iota_{[-,+,+,-,-]}^{[+,-,-,+,+]} = 0.0616$ | 0.7806 |
| | | $\delta_{[+,+,+,-,-]}^{[-,-,-,+,+]} = 0.0758$ | $\iota_{[+,+,+,-,-]}^{[-,-,-,+,+]} = 0.0129$ | 0.9534 |
| | | $\delta_{[-,-,-,+,-]}^{[+,+,+,-,+]} = 0.0000$ | $\iota_{[-,-,-,+,-]}^{[+,+,+,-,+]} = -0.0679$ | 0.7587 |
| | | $\delta_{[+,-,-,+,-]}^{[-,+,+,-,+]} = 0.0758$ | $\iota_{[+,-,-,+,-]}^{[-,+,+,-,+]} = 0.0129$ | 0.9534 |
| | | $\delta_{[-,+,-,+,-]}^{[+,-,+,-,+]} = 0.1364$ | $\iota_{[-,+,-,+,-]}^{[+,-,+,-,+]} = 0.0778$ | 0.7248 |
| | | $\delta_{[+,+,-,+,-]}^{[-,-,+,-,+]} = 0.0000$ | $\iota_{[+,+,-,+,-]}^{[-,-,+,-,+]} = -0.0679$ | 0.7587 |
| | | $\delta_{[-,-,+,+,-]}^{[+,+,-,-,+]} = 0.1364$ | $\iota_{[-,-,+,+,-]}^{[+,+,-,-,+]} = 0.0778$ | 0.7248 |
| | | $\delta_{[+,-,+,+,-]}^{[-,+,-,-,+]} = 0.0758$ | $\iota_{[+,-,+,+,-]}^{[-,+,-,-,+]} = 0.0129$ | 0.9534 |
| | | $\delta_{[-,+,+,+,-]}^{[+,-,-,-,+]} = 0.0455$ | $\iota_{[-,+,+,+,-]}^{[+,-,-,-,+]} = -0.0194$ | 0.9299 |
| | | $\delta_{[+,+,+,+,-]}^{[-,-,-,-,+]} = 0.0152$ | $\iota_{[+,+,+,+,-]}^{[-,-,-,-,+]} = -0.0517$ | 0.8149 |
| | **S1**: 23, 15, 19, 5, 12, 13, 8, 9, 2, 5, 3, 1<br>**S2**: 21, 25, 19, 15, 17, 9, 11, 10, 7, 2, 3, 5<br>**S3**: 10, 3, 5, 2, 9, 11, 10, 15, 17, 25, 21, 23<br>**S4**: 1, 5, 3, 7, 9, 11, 10, 15, 17, 20, 21, 23<br>**S5**: 4, 3, 5, 7, 10, 11, 13, 19, 17, 16, 21, 23 | $\delta_{[-,-,-,-,-]}^{[+,+,+,+,+]} = 0.0156$ | $\iota_{[-,-,-,-,-]}^{[+,+,+,+,+]} = -0.0732$ | 0.7406 |
| | | $\delta_{[+,-,-,-,-]}^{[-,+,+,+,+]} = 0.0313$ | $\iota_{[+,-,-,-,-]}^{[-,+,+,+,+]} = -0.0559$ | 0.8004 |
| | | $\delta_{[-,+,-,-,-]}^{[+,-,+,+,+]} = 0.0938$ | $\iota_{[-,+,-,-,-]}^{[+,-,+,+,+]} = 0.0130$ | 0.9529 |
| | | $\delta_{[+,+,-,-,-]}^{[-,-,+,+,+]} = 0.6563$ | $\iota_{[+,+,-,-,-]}^{[-,-,+,+,+]} = 0.6331$ | 0.0042 |
| | | $\delta_{[-,-,+,-,-]}^{[+,+,-,+,+]} = 0.0000$ | $\iota_{[-,-,+,-,-]}^{[+,+,-,+,+]} = -0.0905$ | 0.6823 |
| | | $\delta_{[+,-,+,-,-]}^{[-,+,-,+,+]} = 0.0313$ | $\iota_{[+,-,+,-,-]}^{[-,+,-,+,+]} = -0.0559$ | 0.8004 |
| | | $\delta_{[-,+,+,-,-]}^{[+,-,-,+,+]} = 0.0000$ | $\iota_{[-,+,+,-,-]}^{[+,-,-,+,+]} = -0.0905$ | 0.6823 |
| | | $\delta_{[+,+,+,-,-]}^{[-,-,-,+,+]} = 0.0782$ | $\iota_{[+,+,+,-,-]}^{[-,-,-,+,+]} = 0.0042$ | 0.9850 |
| | | $\delta_{[-,-,-,+,-]}^{[+,+,+,-,+]} = 0.0000$ | $\iota_{[-,-,-,+,-]}^{[+,+,+,-,+]} = -0.0905$ | 0.6823 |
| | | $\delta_{[+,-,-,+,-]}^{[-,+,+,-,+]} = 0.0000$ | $\iota_{[+,-,-,+,-]}^{[-,+,+,-,+]} = -0.0905$ | 0.6823 |
| | | $\delta_{[-,+,-,+,-]}^{[+,-,+,-,+]} = 0.0313$ | $\iota_{[-,+,-,+,-]}^{[+,-,+,-,+]} = -0.0559$ | 0.8004 |
| | | $\delta_{[+,+,-,+,-]}^{[-,-,+,-,+]} = 0.0000$ | $\iota_{[+,+,-,+,-]}^{[-,-,+,-,+]} = -0.0905$ | 0.6823 |
| | | $\delta_{[-,-,+,+,-]}^{[+,+,-,-,+]} = 0.0313$ | $\iota_{[-,-,+,+,-]}^{[+,+,-,-,+]} = -0.0559$ | 0.8004 |
| | | $\delta_{[+,-,+,+,-]}^{[-,+,-,-,+]} = 0.0313$ | $\iota_{[+,-,+,+,-]}^{[-,+,-,-,+]} = -0.0559$ | 0.8004 |
| | | $\delta_{[-,+,+,+,-]}^{[+,-,-,-,+]} = 0.0000$ | $\iota_{[-,+,+,+,-]}^{[+,-,-,-,+]} = -0.0905$ | 0.6823 |
| | | $\delta_{[+,+,+,+,-]}^{[-,-,-,-,+]} = 0.0000$ | $\iota_{[+,+,+,+,-]}^{[-,-,-,-,+]} = -0.0905$ | 0.6823 |

**Table 4.** Example of *data trend* measurement ($\iota_{[\mp]}^{[\pm]}$) for $H(S1, S2, S3, S4, S5, S6) \in \mathbb{R}^6$ ($r_p = 2/2^N$).

| $N$ ($r_p$) | Series ($n = 12$) | $\delta$ [0, 1] | $\iota$ [−1, 1] | $p$ |
|---|---|---|---|---|
| **6** ($r_p$ = 0.0313) | **S1**: 10, 19, 7, 9, 1, 17, 6, 23, 3, 8, 13, 11<br>**S2**: 9, 5, 17, 15, 23, 7, 3, 1, 21, 19, 11, 13<br>**S3**: 9, 5, 17, 15, 23, 10, 3, 1, 18, 19, 11, 18<br>**S4**: 9, 5, 12, 15, 23, 7, 8, 1, 22, 19, 11, 13<br>**S5**: 9, 10, 17, 15, 20, 7, 3, 1, 21, 19, 11, 13<br>**S6**: 15, 19, 7, 9, 1, 17, 21, 23, 3, 5, 13, 20 | $\delta_{[-,-,-,-,-,-]}^{[+,+,+,+,+,+]} = 0.0154$ | $\iota_{[-,-,-,-,-,-]}^{[+,+,+,+,+,+]} = -0.0290$ | 0.8957 |
| | | $\delta_{[+,-,-,-,-,-]}^{[-,+,+,+,+,+]} = 0.0462$ | $\iota_{[+,-,-,-,-,-]}^{[-,+,+,+,+,+]} = 0.0035$ | 0.9874 |
| | | $\delta_{[-,+,-,-,-,-]}^{[+,-,+,+,+,+]} = 0.0000$ | $\iota_{[-,+,-,-,-,-]}^{[+,-,+,+,+,+]} = -0.0452$ | 0.8379 |
| | | $\delta_{[+,+,-,-,-,-]}^{[-,-,+,+,+,+]} = 0.0000$ | $\iota_{[+,+,-,-,-,-]}^{[-,-,+,+,+,+]} = -0.0452$ | 0.8379 |
| | | $\delta_{[-,-,+,-,-,-]}^{[+,+,-,+,+,+]} = 0.0000$ | $\iota_{[-,-,+,-,-,-]}^{[+,+,-,+,+,+]} = -0.0452$ | 0.8379 |
| | | $\delta_{[+,-,+,-,-,-]}^{[-,+,-,+,+,+]} = 0.0000$ | $\iota_{[+,-,+,-,-,-]}^{[-,+,-,+,+,+]} = -0.0452$ | 0.8379 |
| | | $\delta_{[-,+,+,-,-,-]}^{[+,-,-,+,+,+]} = 0.0000$ | $\iota_{[-,+,+,-,-,-]}^{[+,-,-,+,+,+]} = -0.0452$ | 0.8379 |
| | | $\delta_{[+,+,+,-,-,-]}^{[-,-,-,+,+,+]} = 0.0000$ | $\iota_{[+,+,+,-,-,-]}^{[-,-,-,+,+,+]} = -0.0452$ | 0.8379 |
| | | $\delta_{[-,-,-,+,-,-]}^{[+,+,+,-,+,+]} = 0.0000$ | $\iota_{[-,-,-,+,-,-]}^{[+,+,+,-,+,+]} = -0.0452$ | 0.8379 |
| | | $\delta_{[+,-,-,+,-,-]}^{[-,+,+,-,+,+]} = 0.0000$ | $\iota_{[+,-,-,+,-,-]}^{[-,+,+,-,+,+]} = -0.0452$ | 0.8379 |
| | | $\delta_{[-,+,-,+,-,-]}^{[+,-,+,-,+,+]} = 0.0000$ | $\iota_{[-,+,-,+,-,-]}^{[+,-,+,-,+,+]} = -0.0452$ | 0.8379 |
| | | $\delta_{[+,+,-,+,-,-]}^{[-,-,+,-,+,+]} = 0.0000$ | $\iota_{[+,+,-,+,-,-]}^{[-,-,+,-,+,+]} = -0.0452$ | 0.8379 |
| | | $\delta_{[-,-,+,+,-,-]}^{[+,+,-,-,+,+]} = 0.0000$ | $\iota_{[-,-,+,+,-,-]}^{[+,+,-,-,+,+]} = -0.0452$ | 0.8379 |
| | | $\delta_{[+,-,+,+,-,-]}^{[-,+,-,-,+,+]} = 0.0000$ | $\iota_{[+,-,+,+,-,-]}^{[-,+,-,-,+,+]} = -0.0452$ | 0.8379 |
| | | $\delta_{[-,+,+,+,-,-]}^{[+,-,-,-,+,+]} = 0.0462$ | $\iota_{[-,+,+,+,-,-]}^{[+,-,-,-,+,+]} = 0.0035$ | 0.9874 |
| | | $\delta_{[+,+,+,+,-,-]}^{[-,-,-,-,+,+]} = 0.0000$ | $\iota_{[+,+,+,+,-,-]}^{[-,-,-,-,+,+]} = -0.0452$ | 0.8379 |
| | | $\delta_{[-,-,-,-,+,-]}^{[+,+,+,+,-,+]} = 0.0000$ | $\iota_{[-,-,-,-,+,-]}^{[+,+,+,+,-,+]} = -0.0452$ | 0.8379 |
| | | $\delta_{[+,-,-,-,+,-]}^{[-,+,+,+,-,+]} = 0.0000$ | $\iota_{[+,-,-,-,+,-]}^{[-,+,+,+,-,+]} = -0.0452$ | 0.8379 |
| | | $\delta_{[-,+,-,-,+,-]}^{[+,-,+,+,-,+]} = 0.0154$ | $\iota_{[-,+,-,-,+,-]}^{[+,-,+,+,-,+]} = -0.0290$ | 0.8957 |
| | | $\delta_{[+,+,-,-,+,-]}^{[-,-,+,+,-,+]} = 0.0000$ | $\iota_{[+,+,-,-,+,-]}^{[-,-,+,+,-,+]} = -0.0452$ | 0.8379 |
| | | $\delta_{[-,-,+,-,+,-]}^{[+,+,-,+,-,+]} = 0.0000$ | $\iota_{[-,-,+,-,+,-]}^{[+,+,-,+,-,+]} = -0.0452$ | 0.8379 |
| | | $\delta_{[+,-,+,-,+,-]}^{[-,+,-,+,-,+]} = 0.0000$ | $\iota_{[+,-,+,-,+,-]}^{[-,+,-,+,-,+]} = -0.0452$ | 0.8379 |
| | | $\delta_{[-,+,+,-,+,-]}^{[+,-,-,+,-,+]} = 0.0154$ | $\iota_{[-,+,+,-,+,-]}^{[+,-,-,+,-,+]} = -0.0290$ | 0.8957 |
| | | $\delta_{[+,+,+,-,+,-]}^{[-,-,-,+,-,+]} = 0.0308$ | $\iota_{[+,+,+,-,+,-]}^{[-,-,-,+,-,+]} = -0.0127$ | 0.9541 |
| | | $\delta_{[-,-,-,+,+,-]}^{[+,+,+,-,-,+]} = 0.0000$ | $\iota_{[-,-,-,+,+,-]}^{[+,+,+,-,-,+]} = -0.0452$ | 0.8379 |
| | | $\delta_{[+,-,-,+,+,-]}^{[-,+,+,-,-,+]} = 0.0000$ | $\iota_{[+,-,-,+,+,-]}^{[-,+,+,-,-,+]} = -0.0452$ | 0.8379 |
| | | $\delta_{[-,+,-,+,+,-]}^{[+,-,+,-,-,+]} = 0.0462$ | $\iota_{[-,+,-,+,+,-]}^{[+,-,+,-,-,+]} = 0.0035$ | 0.9874 |
| | | $\delta_{[+,+,-,+,+,-]}^{[-,-,+,-,-,+]} = 0.0000$ | $\iota_{[+,+,-,+,+,-]}^{[-,-,+,-,-,+]} = -0.0452$ | 0.8379 |
| | | $\delta_{[-,-,+,+,+,-]}^{[+,+,-,-,-,+]} = 0.0000$ | $\iota_{[-,-,+,+,+,-]}^{[+,+,-,-,-,+]} = -0.0452$ | 0.8379 |
| | | $\delta_{[+,-,+,+,+,-]}^{[-,+,-,-,-,+]} = 0.0000$ | $\iota_{[+,-,+,+,+,-]}^{[-,+,-,-,-,+]} = -0.0452$ | 0.8379 |
| | | $\delta_{[-,+,+,+,+,-]}^{[+,-,-,-,-,+]} = 0.6615$ | $\iota_{[-,+,+,+,+,-]}^{[+,-,-,-,-,+]} = 0.6504$ | 0.0032 |
| | | $\delta_{[+,+,+,+,+,-]}^{[-,-,-,-,-,+]} = 0.1231$ | $\iota_{[+,+,+,+,+,-]}^{[-,-,-,-,-,+]} = 0.0845$ | 0.7022 |